# New form of the Boltzmann constant from search for a permanent electric dipole moment of K, Rb or Cs atom

Pei-Lin You

Institute of Quantum Electronics, Guangdong Ocean University, zhanjiang 524025, China.

Using special capacitors our experiments discovered that the ground state K, Rb or Cs atom is polar atom with a large permanent electric dipole moment (EDM) of the order of $ea_o$ ($a_o$ is Bohr radius) as excited state of hydrogen atom. But we can not calculate the EDM of an atom using Boltzmann constant. The mechanism of polar atoms by which orientation polarization arises completely differs from polar molecules. The orientation polarization of polar molecule, such as HCl or $H_2O$ etc, is a molecule as a whole turned toward the direction of an external field. Unlike polar molecules, the orientation polarization of polar atom, such as K, Rb or Cs atom, is only the valence electron in the outermost shell, along a highly elliptical orbit, turned toward the direction of the field, but the rest of the atom does not move! The article derived the new form of Boltzmann constant. When it was applied to orientation polarization of polar atoms, its value $k \ll 1.38 \times 10^{-23}$ J/K(for example, $k=7.14 \times 10^{-29}$ J/K for Cs atom) . When it was applied to any molecule, non-polar atom or translation and vibration of polar atom, its value $k=1.38 \times 10^{-23}$ J/K. The reason is that the nucleus moves together with the rest of the molecule or atom. ( see arXiv: 0809.4767; 0810.0770; 0810.2026)

**PACS:**  32.10.Dk, 11.30.Er, 32.60.+i , 03.50.De

**1. Introduction**   New information from upcoming experiments, such as those at the Large Hadron Collider (LHC) at CERN, should help us find answers to many outstanding questions haunting the field of particle physics for few decades. Non-accelerator probes like the search for the permanent electric dipole moment (EDM) of any atom played a crucial role in attempting to establish new physics beyond the Standard Model. However, quantum mechanics thinks that atoms do not have EDM because of their spherical symmetry, or at most has an extremely small EDM ($d_{atom} \leq e \times 10^{-20}$cm) that might arise from weak interactions[1]. Therefore, there is no polar atom in nature except for polar molecules. **This specious viewpoint led worldwide physicists away from the right path for many decades**[2].

Search for an EDM of atom, the most sensitive of which is done with $^{199}$Hg(the result is $d(Hg)=-[1.06 \pm 0.49 \text{(stat)} \pm 0.40 \text{(syst)}] \times 10^{-28}$e.cm )[3]. Experiments to search for an EDM of atom for more than 50 years, not only no large EDM but also no conclusive results have been obtained so far ($d_{atom} \ll 10^{-24}$ e.cm)[3-7].

The linear Stark effect shows that the first excited state of hydrogen atom has very large EDM, $d_H = 3e\, a_o = 1.59 \times 10^{-8}$e.cm ($a_o = 0.53 \times 10^{-8}$ cm is Bohr radius) [8,9]. The radius of the hydrogen atom is $r_H = 4a_o = 2.12 \times 10^{-8}$ cm, it is almost the same as the $^{199}$Hg ($r_{Hg} = 1.51 \times 10^{-8}$ cm) [10], but the discrepancy between their EDM is by some twenty orders of magnitude! How do explain this inconceivable discrepancy? The existing theory can not answer the problem! The existing theory thinks that in quantum mechanics there is no such concept as the path of an electron[8]. No one will give you any deeper explain of the inconceivable discrepancy! However, a hydrogen atom ($n \geq 2$) has a nonzero EDM in the semi-classical theory of atom. The electron in a hydrogen atom ($n \geq 2$) moves along a quantization elliptic orbit. We can draw a straight line perpendicular to the major axis of the elliptic orbit through the nucleus in the orbital plane. The straight line divides the elliptic orbit into two parts that are different in size. According conservation of the area velocity, the average distance between the moving electron and the static nucleus is larger and the electron remains in the large part longer than in the small part. As a result, the time-averaged value of the electric dipole moment over a period is nonzero.

We think that the only one valence electron of alkali atoms moves along a highly elliptical orbit, the so-called diving orbits, as the excited state of the hydrogen atom[11]. So we conjecture that the ground state alkali atoms, as the K, Rb and Cs atoms, may have large EDM of the order of e $a_o$.

In all experiments, they measured microcosmic Larmor precession frequency of individual atom based on weak interactions. We have submitted three articles on the similar topic, however, with measuring macroscopic electric susceptibility ($x_e$) of K, Rb or Cs vapor containing a large number of atoms based on electromagnetic interactions. Our experiments showed that a ground state neutral K, Rb or Cs atom is polar atom with a large EDM of the order of e $a_o$ as excited state of hydrogen atom[12-14]. These results are the product of eight years of intense research[15]. Few experiments in atomic physics have produced a result as surprising as this one !

In order for an atom or elementary particle to possess a EDM, time reversal (T) symmetry must be violated, and through the CPT theorem CP(charge conjugation and parity) must be violated as well[1]. The first example of CP violation was discovered in 1964, but it has been observed only in the $K_o$ mesons. After 38 years, the BaBar collaboration at SLAC announced the second example. "The results gave clear evidence for CP violation in B mesons. However, the degree of CP violation now confirmed is not enough on its own to account for the matter-antimatter imbalance in the Universe." "In the past few years we have learned to cleanly isolate one event in a million." (SLAC Press Release July 23, 2002 and Sep 28, 2006).Theorists found it hard to see why CP symmetry should be broken at all and even harder to understand why any imperfection should be so small[16]. This fact suggests that there must be other ways in which CP symmetry breaks down and more subtle effects must be examined. So EDM experiments are now considered an ideal probe for evidence of new sources of CP violation. If an EDM is found, it will be compelling evidence for the existence of new sources of CP violation. **Our experiments showed that new example of CP violation occurred in K, Rb or Cs atoms and it is a classic example of how understanding of our Universe advances through atomic physics research.**

In the following two sections we shall use the two different methods to treat the polar atom (for instance, Cs atoms), and polar molecules (for instance, $H_2O$ molecules). We shall see the true significance of Boltzmann constant and the difference and relation between the two methods.

**2. Orientation polarization of polar molecules**   Many molecules do have *permanent* electric dipole moments. This molecule is called a polar molecule, such as $H_2O$, HCl, etc. A polar molecule tends to be oriented parallel to the applied electric filed because of the torque it experiences. The molecular electric dipole moments are of the order of magnitude of the electronic charge ($1.6\times10^{-19}$ C) multiplied by the molecular dimensions ($10^{-10}$ m), or about $10^{-30}$ coulomb-meters. The electric susceptibility of polar gases should be of the order of $x_e \leqslant 10^{-3}$, typical values are 0.0126 for water vapor, 0.0046 for HCl gas, etc.

We considered the orientation polarization of polar molecules in a gas by using the methods of statistical mechanics. The local field acting on a molecule in a gas is almost the same as the external field **E**. All molecules are assumed to possess a EDM $\mathbf{d_o}$. In the absence of a field thermal agitation keeps the molecules randomly oriented so that there is no net dipole moment. After switching on a field **E**, a molecule with dipole moment acquires an additional potential energy $U = -d_o E \cos\theta$, where $\theta$ is the angle between $\mathbf{d_o}$ and **E**. Because of the collisions of the molecules in their thermal motion, the probability distribution at thermal equilibrium is now modified by the presence of a Boltzmann factor $\exp(-U/kT)$. Consequently there will be an average dipole moment. In addition, all the molecules also acquire an induced dipole moment. If the molecular polarizability is $\alpha$, there is an additional induced polarization $N\alpha\varepsilon_o E$, where N is the number density of molecules. Finally, we have the result that the polarization, the average dipole moment per unit volume, is

$$P = N\alpha\varepsilon_o E + N d_o L(a) \qquad (1)$$

The electric susceptibility of a gaseous polar dielectric is[17]

$$x_e = N\alpha + N d_o L(a) / \varepsilon_o E \qquad (2)$$

where $a = d_o E /kT$, k is Boltzmann constant, $\varepsilon_o$ is the permittivity of free space, L(a) is called the Langevin function

$$L(a) = [(e^a + e^{-a}) / (e^a - e^{-a})] - 1/a \qquad (3)$$

The Langevin function L(a) is equal to the mean value of $\cos\theta$ [17]:

$$<\cos\theta> = \mu \int_0^\pi \cos\theta \exp(d_o E \cos\theta /kT) \sin\theta \, d\theta = L(a), \quad \mu = [\int_0^\pi \exp(d_o E \cos\theta /kT) \sin\theta \, d\theta]^{-1} \qquad (4)$$

where μ is a normalized constant. This result shows that L(a) is the percentage of polar molecule lined up with the field in the total number. When a<<1 and L(a)≈a/3, when a>>1 and L(a)≈1. For all polar molecules $d_oE/kT$ is normally very small, so L(a)≈a/3= $d_oE/3kT$, we have

$$x_e = N\alpha + Nd_o^2/3k\varepsilon_o T = A+B/T. \tag{5}$$

where A and the slope B is constant at a fixed density. Note that the electric susceptibility caused by the orientation of polar molecules is inversely proportional to the absolute temperature: $x_e = B/T$ while the induced electric susceptibility due to the distortion of electronic motion in atoms or molecules is temperature independent: $x_e=A$. Clearly, this difference in temperature dependence offers a means of separating the polar and non-polar substances experimentally. Because of $x_e=C/C_o-1$, where $C_o$ is the vacuum capacitance and C is the capacitance of the capacitor filled with the material, by measuring the capacitance C at different temperature T, it is possible to distinguish between permanent and induced dipole moment.

The electric susceptibility caused by the orientation of gaseous polar molecules is given by the expression

$$x_e = Nd_o^2/3kT\varepsilon_o \tag{6}$$

The slope is B= $Nd_o^2/3k\varepsilon_o$ and the EDM of a molecule is

$$d_o = (3k\varepsilon_oB/N)^{1/2} \tag{7}$$

R.P. Feynman considered the orientation polarization of water vapor molecules [8]. He plotted the straight line from four experimental points. Table 1 gives the four experimental data of water vapor at various temperatures [9]. By using least-square method we obtain

$$x_e = A+B/T \approx 0.000178 + 1.50/T \tag{8}$$

From the ideal gas law, the average density of water vapor is $N = P/kT = (1.392\pm0.002)\times10^{25}$ m$^{-3}$. Substituting the slope B =(1.50 ±0.04)K and we work out the EDM of a water molecule

$$d_{H2O} = (3k\varepsilon_oB/N)^{1/2} = (6.28\pm0.18)\times10^{-30} \text{C.m} \tag{9}$$

The result in agreement with the recognized value $d_{H2O}= 6.20\times10^{-30}$ C.m [20].

**Table 1 Experimental measurements of the susceptibility of water vapor at various temperatures**

| T(K) | Pressure(cm Hg) | $x_e$ |
|---|---|---|
| 393 | 56.49 | 0.004002 |
| 423 | 60.93 | 0.003717 |
| 453 | 65.34 | 0.003488 |
| 483 | 69.75 | 0.003287 |

## 3. Orientation polarization of polar atoms

**A. Experimental method and result of Cs atom** We will discuss the orientation polarization of polar atoms with the example—Cs atoms[12]. The first experiment : investigation of the relationship between the capacitance C' of Cs or Hg vapor and their density N. The experimental apparatus were two closed glass containers containing Cs or Hg vapor. Two silver layers **a** and **b** build up the glass cylindrical capacitor(Fig.1,where the plate area $S_1=(5.8\pm0.1)\times10^{-2}$ m$^2$, the plate separation $H_1=(9.6\pm0.1)$mm). The radiuses of the layers **a** and **b** are shown respectively by r and R. Since R-r=H+2△<< r (△=1.2mm), the capacitor can be approximately regarded as a parallel-plane capacitor. The capacitance was measured by a digital meter. The precision of the meter was 0.1pF, the accuracy was 0.5% and the surveying voltage was V=(1.0±0.01)volt. It means that the applied field E=V/$H_1$≈1.1×10$^2$v/m is weak using the meter. This capacitor is equivalently connected in series by two capacitors. One is called C' and contains the Cs or Hg vapor of thickness H , another one is called C" and contains the glass medium of thickness 2△. The total capacitance C can be written as C=C'C"/ (C'+C") or C'=CC"/ (C"−C), where C" and C can be directly measured. When the two containers are empty, they are pumped to a vacuum pressure P≤10$^{-6}$ Pa for 20 hours. The measured capacitances are nearly the same: C'$_{10}$=(54.0±0.1)pF(for Cs) and C'$_{20}$=(52.8±0.1)pF(for Hg). Next, a small amount of the Cs or Hg material with high purity is put in the two containers respectively in a vacuum environment. The two containers are again pumped to vacuum pressure P ≤10$^{-6}$ Pa , then they are sealed. Now, their capacitances are $C'_1=(102.2\pm0.1)$pF and $C'_2=(53.6\pm0.1)$pF respectively at room temperature. We put the two capacitors into a temperature-control stove, raise the temperature of the stove very slow and keep at $T_1=(473\pm0.5)$K for 3 hours. It means that the readings of capacitance are obtained under the condition of Cs or Hg saturated vapor pressure. Two comparable experimental curves are shown in Fig.2. From Ref.[10], the saturated pressure of Hg vapor is $P_{Hg}=2304.4$ Pa at 473K and its capacitance is $C'_{2t}=(56.4\pm0.1)$ pF. From the ideal gas law, the density of Hg vapor is $N_{Hg}=P_{Hg}/kT_1=3.53\times10^{23}$ m$^{-3}$. The formula of saturated vapor pressure of Cs vapor is **P=10$^{6.949-3833.7/T}$** psi and the effective range is 473K $\leqslant$T $\leqslant$623K$^{[10]}$. We obtain the saturated pressure of Cs vapor $P_{Cs}=0.0698$ psi $=481.3$Pa at 473K and the capacitance $C'_{1t}=(3944\pm10)$ pF. The density of Cs atoms is $N_{Cs}=N_1=P_{Cs}/kT_1=7.37\times10^{22}$ m$^{-3}$. The experiment shows that the number density $N_{Hg}$ of Hg vapor is 4.79 times as $N_{Cs}$ of Cs vapor but the capacitance change($C'_{2t}-C'_{20}$) of Hg vapor being only 1/1080 of ($C'_{1t}-C'_{10}$) of Cs vapor! So unlike Cs atom, a Hg atom is non-polar atom.

The second experiment: investigation of the relationship between the electric susceptibility($\chi_e$) of Cs or Hg vapor and temperatures(T) under each fixed density. The experimental apparatus was a closed glass container containing Cs vapor. Two stainless steel tubes **a** and **b** build up the glass cylindrical capacitor (Fig.3). Since R-r=H<<r, the capacitor could be approximately regarded as a parallel-plane capacitor. The capacitance was still measured by the digital meter, and vacuum capacitance $C_0=(66.0\pm0.1)$pF(where $S_2=(5.7\pm0.1)\times10^{-2}$ m$^2$, $H_2=(7.6\pm0.1)$mm ). Then the capacitor filled with Cs vapor at the fixed density $N_2$ and was put into the stove. By measuring the electric susceptibility $\chi_e$ of Cs vapor at different temperature T, we obtain $\chi_e=A+B/T$ $\approx$B/T, where the slope B=(320$\pm$4) K and A$\approx$0 . The capacitance of an identical capacitor containing Hg vapor was measured and we found that the slope is nearly zero, B$\approx$0.0K( $\chi_e$<0.07 is nearly constant). The experimental results are shown in Fig.4.

The third experiment: measuring the capacitance of Cs vapor at various voltages (V) under a fixed $N_2$ and $T_2$. The apparatus was the same as the second experiment ($C_0=(66.0\pm0.1)$pF) and the method is shown in Fig.5. C was the capacitor filled with Cs vapor to be measured, which was kept at $T_2=(323\pm0.5)$K, and $C_d$ was used as a standard capacitor. Two signals $V_c(t)=V_{co}\cos\omega t$ and $V_s(t)=V_{so}\cos\omega t$ were measured by a digital oscilloscope having two lines. The two signals had the same frequency and always the same phase at different voltages when the frequency was higher than a certain value. It indicates that capacitor C filled with Cs vapor was the same as $C_d$ , a pure capacitor without loss. From Fig.5, we have $C=C_d(V_{so}-V_{co})/V_{co}$. In the experiment $V_{so}$ can be adjusted from zero to 800V. The capacitance C at various voltages was shown in Fig.6. When $V_{co}=V_1=(0.3\pm0.02)$volt, $C_1=(130.0\pm0.4)$pF is approximately constant, and $\chi_e=C_1/C_o-1=0.9697$. With the increase of voltage, the capacitance decreases gradually. When $V_{co}=V_2=(560\pm2)$V, $C_2=(68.0\pm0.4)$ pF, it approaches saturation and $\chi_e=C_2/C_0-1\leqslant0.0303\approx0$. If nearly all dipoles in a gas turn toward the direction of the field, this effect is called the saturation polarization. The C-V curve shows that the saturation polarization of Cs vapor was obvious when E= $V_2/H_2\geqslant7.4\times10^4$v/m.

B. **Polarization theory of polar atoms** The electric susceptibility of gaseous polar molecules is
$$\chi_e=N\alpha+N d_o L(a)/\varepsilon_o E \tag{2}$$

Next we will consider how this equation is applied to Cs atoms. The induced dipole moment of an Cs atom is $d_{int}=G\varepsilon_o E$, where $G=59.6\times10^{-30}$m$^3$ $^{[21]}$. Due to the most field intensity is E$\leqslant7.4\times10^4$v/m in the experiment, then $d_{int}\leqslant3.9\times10^{-35}$ C.m $=2.4\times10^{-14}$e.cm can be neglected. From Eq.(2) we obtain
$$\chi_e=N d L(a)/\varepsilon_o E \tag{10}$$

where d is the EDM of an Cs atom and N is the number density of Cs vapor. **L(a)= < cos $\theta$ > is the percentage of polar atoms which lined up with the field in the total number .** Note that $\chi_e=\varepsilon_r-1=C/C_o-1$, where $\varepsilon_r$ is the dielectric constant , E=V/H and $\varepsilon_o=C_o H/S$, leading to **the polarization equation of Cs atoms**
$$C-C_o=\beta L(a)/a, \tag{11}$$

where $\beta=SNd^2/kTH$ is a constant. Due to a=dE/kT= dV/kTH **we obtain the first formula of atomic EDM**
$$d_{atom}=(C-C_o)V/L(a)SN \tag{12}$$

In order to work out **L(a)** and **a** of the first experiment, note that in the third experiment when the field is weak ($V_1=0.3$ v), $a_1 \ll 1$ and $L(a_1) \approx a_1/3$, $C_1 - C_o = \beta/3$ and $\beta = 192$ pF. When the field is strong($V_2= 560$ v), $a_2 \gg 1$ and $L(a_2) \approx 1$, $C_2 - C_o = L(a_2)\beta/a_2 \approx \beta/a_2$. We work out $a_2 \approx \beta/(C_2 - C_o)=96$, $L(a_2) \approx L(96)=0.9896$, $a_2 = \beta L(a_2)/(C_2 - C_o)=95$, $L(a_2)=L(95)=0.9895$. Due to $\mathbf{a}=dE/kT=dV/kTH$, so $\mathbf{a}/a_2 = VT_2H_2/T_1H_1V_2$. Substituting the corresponding values, we obtain $\mathbf{a}=0.0917$ and $L(\mathbf{a}) \approx \mathbf{a}/3=0.0306$. $L(\mathbf{a}) \approx 0.0306$ means that only 3.06％ of Cs atoms have be oriented in the direction of the field when the field $E=V/H_1=1.1 \times 10^2$V/m. Substituting the values: $S_1=5.8 \times 10^{-2}$ m$^2$, $N_1 =7.37 \times 10^{22}$ m$^{-3}$, $V=1.0$volt, $C - C_o = C'_{1t} - C'_{10} = 3890$pF and L(a), we work out

$$d_{Cs} =(C - C_o)V / L(a) S_1 N_1 = 2.97 \times 10^{-29} \text{C.m} = 1.86 \times 10^{-8} \text{e.cm} \approx 3.5\ e\ a_o \qquad (13)$$

The statistical errors is $\triangle d_1/d = \triangle C/C + \triangle C_o/C_o + \triangle S_1/S_1 + \triangle V/V + \triangle N_1/N_1 < 0.12$, considering all systematic error $\triangle d_2/d < 0.08$ (including $\triangle L(a)/L(a)$), and the combination error $\triangle d/d < 0.15$. We find that

$$\mathbf{d(Cs)=[2.97 \pm 0.36(stat) \pm 0.24(syst)] \times 10^{-29} \text{C.m} = [1.86 \pm 0.22(stat) \pm 0.15(syst)] \times 10^{-8} \text{e. cm}} \qquad (14)$$

**Although above calculation is simple, but no physicist completed the calculation up to now!**

### 4. Discussion

①The formula $d_{atom} =(C - C_o)V/L(a)SN$ can be justified easily. The magnitude of the dipole moment of an atom is $d = e\ r$. N is the number of atoms per unit volume. L(a) is the percentage of Cs atoms lined up with the field in the total number. Suppose that the plates of the capacitor have an area S and separated by a distance H, the volume of the capacitor is SH. When a field is applied, the Cs atoms tend to orient in the direction of the field. On the one hand, the change of the charge of the capacitor is $\triangle Q=(C-C_0)V$. On the other hand, when the Cs atoms are polarized by the orientation, the total number of Cs atoms lined up with the field is $SHNL(a)$. The number of layers of Cs atoms which lined up with the filed is $H/r$. Because the positive and negative charges cancel out each other inside the Cs vapor, the polarization only gives rise to a net positive charge on one side of the capacitor and a net negative charge on the opposite side. Then the change $\triangle Q$ of the charge of the capacitor is $\triangle Q=SHN L(a)e / (H/r)= SN L(a)d = (C - C_0)V$, so the EDM of an Cs atom is $\mathbf{d = (C - C_0)V/ SN L(a).}$

②If Cs atom is the polar atom, we should calculate the EDM of an Cs atom using the same method such as water molecules from $d=(3k\varepsilon_o B / N_2)^{1/2}$. Note that $a_1 = a_2 V_1/V_2 =0.0509$, the polarization is respectively

$$P_1 = \varepsilon_o x_e E_1 =(C_1 - C_o)V_1/ S_2 = N_2 d\ L(a_1)= N_2 d\ a_1/3=0.017\ N_2 d \approx N_2 d/58.9 \qquad (15)$$

$$P_2 = \varepsilon_o x_e E_2 =(C_2 - C_o)V_2/ S_2 = N_2 d\ L(a_2) =0.9895 N_2 d \approx N_2 d \qquad (16)$$

Due to the density $N_2$ is unknown in the second and the third experiments, from Eq.(9) and Eq.(14) we can obtain

$$N_2 = (C_2 - C_o)V_2 L(a) S_1 N_1 / (C - C_o)V L(a_2) S_2 = 6.70 \times 10^{20} \text{m}^{-3} \qquad (17)$$

From $d=(3k\varepsilon_o B / N_2)^{1/2}$, note that $k= dE/aT_1= dV/aT_1H_1$, **we obtain the second formula of atomic EDM**

$$\mathbf{d(Cs)=(3k\varepsilon_o B/ N_2)^{1/2}= 3\ V\ \varepsilon_o B/ aT_1H_1N_2 = 3.04 \times 10^{-29} \text{C.m}= 1.90 \times 10^{-8} \text{e.cm}} \qquad (18)$$

**Using two different methods and different experimental data we obtain the same result, it proved that the data are reliable and the EDM of an Cs atom has been measured accurately.**

③If Cs atom has a large EDM, why has not been observed in other experiments? This is an interesting question. In Eq.(11) let the new function

$$f(a)= L(a)/a= [(e^a + e^{-a})/a(e^a - e^{-a})] - 1/a^2 \qquad (19)$$

From $f''(a)=0$, we work out the knee of the function $L(a)/a$ at

$$a=1.9296813 \approx 1.93 \qquad (20)$$

Corresponding knee voltages $V_k=11.4$V and the knee field $E_k=V_k/H_2 \approx 1.5 \times 10^3$V/m. By contrast with the curve in Fig.6, it is clear that our polarization equation is valid. The curve shows clearly that only under the very weak field ($E<E_k$ i.e. $E<1.5 \times 10^3$V/m), the large EDM of Cs atom can be observed. **Regrettably, nearly all scientists in this field disregard the very important problem.**

④When the saturation polarization appeared, due to the angle $\theta = 0$, then $<\cos\theta>= L(a) \approx 1$ and this will happen only if $a \gg 1$[17]. Because the most EDM of polar molecules $d_o$ is roughly $10^{-29}$ C.m and the breakdown field intensity of gaseous dielectric is roughly $10^7$v/m. The most potential energy $d_oE$ of a molecule is about $6.3 \times 10^{-4}$ ev. The average kinetic energy $kT$ of each molecule in a gas is about 0.03ev at ordinary temperatures (T=300k), and most value $a_{max}=d_oE/kT <0.03 \ll 1$. It means that the saturation polarization can happen only if temperatures near absolute zero for all polar molecules. So no scientist has observed the saturation polarization of any gaseous dielectric till now. R.P. Feynman stated that " *when a filed is applied, if all the dipoles in a gas were to line up, there would be a very large polarization, but that does not happen*" [18]. The saturation polarization of K, Rb or Cs vapor in ordinary temperatures is an entirely unexpected discovery.

⑤ The saturation polarization of K, Rb or Cs vapor can not be explained by existing theories and it tells us that the mechanism of polar atoms by which orientation polarization arises completely differs from polar molecules. The orientation polarization of the polar molecule, such as HCl or $H_2O$ etc, is the molecule as a whole turned toward the direction of an external field. We can measure the average rotation energy of individual molecule by using kT. So we can work out the probability distribution at thermal equilibrium using the Boltzmann factor exp (–U/kT). **Unlike polar molecules, the orientation polarization of the polar atom, such as K, Rb or Cs atom, is only the valence electron in the outermost shell turned toward the direction of the field, but the rest of the atom(include the nucleus) does not move ! Due to the rotational inertia of an electron is much less than that of an atom, when finding the orientation probability distribution of polar atoms at thermal equilibrium, we should use the new Boltzmann constant *k* instead of k=1.38×$10^{-23}$ J/K.** Its form is

$$k = dE/aT \quad \text{or} \quad k = U/aT \tag{21}$$

where U is the potential energy of atoms. We will calculate the average rotation energy of polar atoms by dE/a. Table 2 gives three values of new Boltzmann constant for Cs atom, where d= 2.974×$10^{-29}$ C.m.

**Table 2 The three values of new form of Boltzmann constant for Cs atom**

| T(K) | V | H | a | E | k |
|---|---|---|---|---|---|
| 323 | 0.3 | 7.6 mm | 0.0509 | 39.5 V/m | 7.14×$10^{-29}$ J/K |
| 323 | 560 | 7.6 mm | 95 | 7.37×$10^4$ V/m | 7.14×$10^{-29}$ J/K |
| 473 | 1.0 | 9.6 mm | 0.09171 | 104.2 V/m | 7.14×$10^{-29}$ J/K |

**Due to *k*=7.14×$10^{-29}$ J/K<< 1.38×$10^{-23}$ J/K, so the saturation polarization of Cs atoms is easily observed. From k= dE/aT= dV/aTH, the second formula of atomic EDM is $d_{atom}$=(3k $\varepsilon_o$B / N)$^{1/2}$ =3V $\varepsilon_o$B/aTHN.**

⑥We will calculate new value of Boltzmann constant for water molecule.

Due to for all polar molecules a<<1( like water molecule, when V=560volt H=7.6mm T=393K, a=0.00000064), so a=3L(a)=3 (C - $C_0$ )V/dSN, and E=V/H, we get

$$k = dE/aT = S Nd^2 /3 (C - C_0) H T \tag{22}$$

From Eq.(8), $\chi_e$'= $\chi_e$– A = B/T, using $\varepsilon_o$= $C_0$H /S, N = P/ *k* T, $\chi_e$=(C - $C_0$)/$C_0$ and B= $\chi_e$' T, we obtain

$$k = dE/aT = d (P/3BT \varepsilon_o)^{1/2} \tag{23}$$

where d= 6.28×$10^{-30}$ C.m is the experimental value[20]. Table 4 gives four values of new Boltzmann constant.

**Table 4 The four values of new form of Boltzmann constant for water molecule**

| T(K) | Pressure(cm Hg) | $\chi_e$' | B | (P/3BT $\varepsilon_o$)$^{1/2}$ | k= d (P/3BT $\varepsilon_o$)$^{1/2}$ |
|---|---|---|---|---|---|
| 393 | 56.49 | 0.003824 | 1.503 | 0.2191×$10^7$ | 1.376×$10^{-23}$ J/K |
| 423 | 60.93 | 0.003539 | 1.497 | 0.2198×$10^7$ | 1.380×$10^{-23}$ J/K |
| 453 | 65.34 | 0.003310 | 1.499 | 0.2198×$10^7$ | 1.380×$10^{-23}$ J/K |
| 483 | 69.75 | 0.003109 | 1.502 | 0.2197×$10^7$ | 1.380×$10^{-23}$ J/K |

When the new form of Boltzmann constant was applied to any molecule, non-polar atom or translation and vibration of polar atom, its value k=1.38×$10^{-23}$ J/K. The reason is that the nucleus moves together with the rest of the molecule or atom. When it was applied to orientation polarization of polar atoms, its value k<< 1.38×$10^{-23}$ J/K. **Its success proved that the true significance of Boltzmann constant is *k*= U/aT rather than k=1.38×$10^{-23}$ J/K.** Above phenomenon exceeded all expert's expectation and can not be explained by existing theories. **Although the experiments are simple, but no physicist completed the experiments up to now!**

Our measuring process and experimental results can be easily repeated in any laboratory because of the details of the experiments are described in the three article. Our experimental apparatus are still kept, we welcome anyone who is interested in the experiments to visit and examine it.

**Acnowledgement**   The authors thank to Xiang-You Huang, Ri-Zhang Hu, Zhao Tang , Rui-Hua Zhou, Shao-Wei Peng, Yong Chen, Xue-ming Yi, Xiao-ming Wang, Xing Huang, and Engineer Jia You for their help in the work.

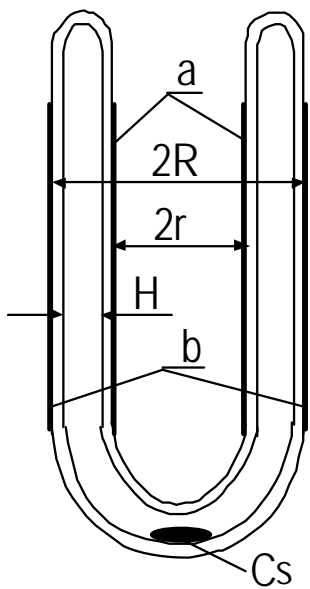

**Fig.1** This is the longitudinal section of the first experimental apparatus. Silver layers **a** and **b** build up a cylindrical glass capacitor. (not to scale).

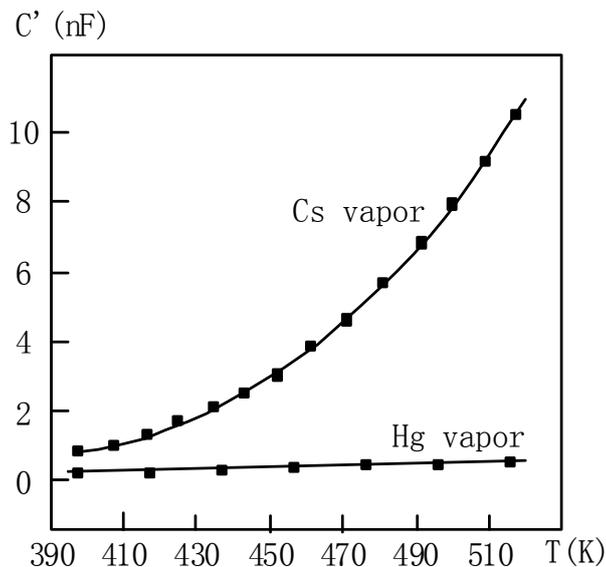

**Fig.2** Two curves showed that the relationship between the capacitance C' of Cs or Hg vapor and the density N respectively, where 1nF=$10^3$ pF.

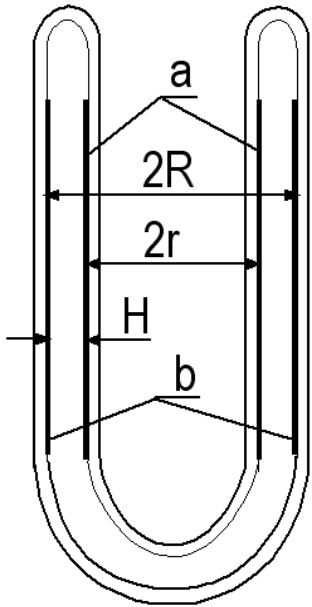

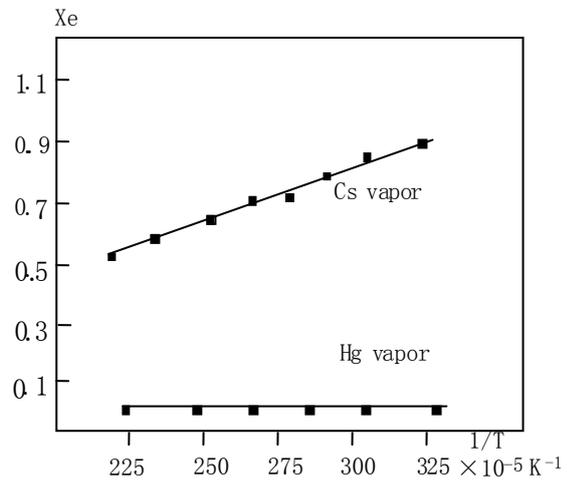

**Fig.3** This is the longitudinal section of another two experimental apparatus. Two round stainless steel tubes **a** and **b** build up a glass capacitor (not to scale).

**Fig.4** The temperature (T) dependence of the susceptibility ($x_e$) of Cs or Hg vapor. The slope of Cs vapor is B≈320(k) but the slope of Hg vapor is nearly zero, B≈0.0(k).

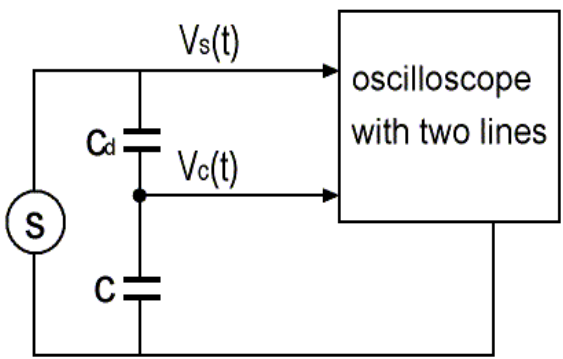

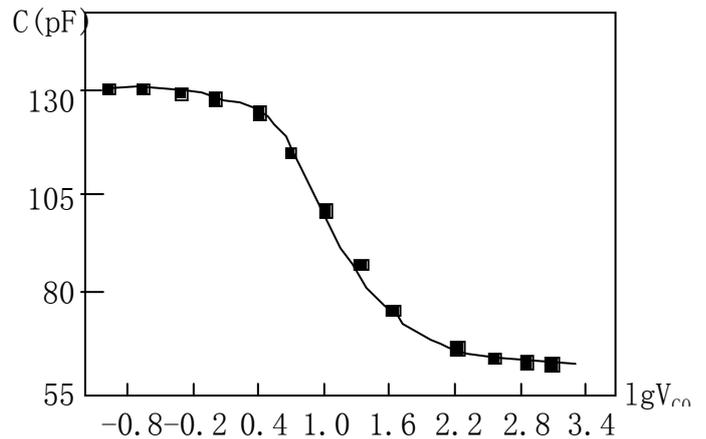

**Fig.5** The diagram shows the experimental method, in which C is capacitor filled with Cs vapor to be measured and Cd is a standard one, where $V_s(t) = V_{so} \cos \omega t$ and $V_c(t) = V_{co} \cos \omega t$.

**Fig.6** The experimental curve shows that the saturation polarization effect of the Cs vapor is obvious when E≥$7.4×10^4$v/m.

8